\documentstyle[12pt]{article}
\title{{\normalsize
\begin{flushright}SUSX-TH-95/72\\
{\tt hep-ph/9505357}\\
May 1995\\
\end{flushright}}
\vspace{1 cm}
Instabilities of Electroweak Strings}
\author{Michael Goodband\thanks{e-mail:
{\tt m.j.goodband@sussex.ac.uk }}
and Mark Hindmarsh\thanks{e-mail:
{\tt m.b.hindmarsh@sussex.ac.uk }}
}
\date{}

\begin{document}
\maketitle
\vspace{-24pt}
\begin{center}
{\it
 School of Mathematical and Physical Sciences\\
University of Sussex\\
Brighton BN1 9QH\\
UK
}
\end{center}

\vfill

\begin{abstract}

We investigate the instabilities of low winding number electroweak
strings using standard numerical techniques of linear algebra.
For strings of unit winding we are able to confirm and extend
existing calculations of the unstable region in the $(m_H /m_W ,
\sin^2 \theta_W )$ plane. For strings of higher winding number
we map the unstable regions for the various decay modes.

\end{abstract}
\newpage

It has been known for some time that certain field theory models with
spontaneous symmetry breaking contain toplogical defects, such as
domain walls, strings and monpoles. These defects are stable because
their field configurations wrap around the vacuum manifold, and to
unwind them requires lifting the Higgs field to its vacuum value in
the unbroken phase,throughout an infinite volume of space. There is
therefore an infinite energy barrier separating these defect solutions
from the vacuum.

Recently, defect solutions have been found in theories where they are
not topologically stable \cite{V&A,Hind,Vach}. These solutions may
be a maximum or a local minimum
of the energy, depending upon the dynamics of the model in question.
Such `embedded defects' have been found in the electroweak model:
$SU(2)_L \times U(1)_Y \rightarrow U(1)_{em}$. This contains
two embedded $U(1) \rightarrow I$ symmetry breakings which can
both give rise to  Nielsen-Olesen vortex solutions \cite{Vach}.
These solutions are called $\tau$- and $Z$-strings depending
upon how one chooses the embedding.
$\tau$-strings are expected to be unstable for all values of the
parameters \cite{T&B} and so we will only consider $Z$-strings.

The stability of $Z$-strings for a range of parameters was addressed in
James {\it et al} \cite{James} where it was shown that
for physical values of the parameters
the $N=1$ string is unstable. The stability analysis did not extend
down to small values of the Higgs self coupling because of numerical
problems caused by the widely different scales of the two cores of
the string. The method adopted for fixing the
gauge, locating the unstable mode and computing the eigenvalues was
not wholly transparent either.

In Perkins \cite{Perk} and Ach\'ucarro {\it et al}
\cite{Achu} instabilities due to $W$-condensation were
investigated, and in Ach\'uccaro {\it et al} it was shown that a
$W$-condensed string of winding number $N$ was equivalent to a string
of winding number $N-1$. In addition to this decay mode Ach\'uccaro
{\it et al} also showed that there is another distinct physical decay
mode for large $N$ strings, where the upper component of the Higgs field
acquires a non-zero value at the string core. This decay mode should
result in the complete unwinding of the string.

MacDowell and T\"ornkvist \cite{D&T} showed that for some, unspecified,
regions of the parameter space a string of winding number $N$ is
unstable to the formation of $W$-fields with angular momentum $m$
such that $-2N < m <0$.

In this letter we seek to complete the stability line of James
{\it et al}, investigate the two decay modes of Ach\'ucarro {\it et al}
and investigate the $2N-1$ decay modes of MacDowell and T\"ornkvist.
We are careful to fix the gauge at the outset, and we use more usual
eigenvalue methods than were adopted in James {\it et al}. We do not,
however, follow Ach\'ucarro {\it et al} in tracking the full
non-linear evolution of the fields.

We consider the bosonic sector of the Weinberg-Salam model with the
lagrangian
$$
{\cal L} = -\frac{1}{4} B_{\mu \nu} B^{\mu \nu}
-\frac{1}{4} G_{\mu \nu}^a G^{a\mu \nu}
+ (D_{\mu} \phi )^{\dagger} (D^{\mu} \phi )
- \lambda (\phi^{\dagger} \phi - \eta^2 /2 )^2,
$$
where
\begin{eqnarray}
D_{\mu} & = & (\partial_{\mu} -ig\sigma^a W_{\mu}^a /2
 -ig^{\prime} B_{\mu} /2 )\phi \nonumber \\
B_{\mu \nu} & = & \partial_{\mu} B_{\nu} - \partial_{\nu} B_{\mu}
\nonumber \\
G_{\mu \nu}^a & = & \partial_{\mu} W_{\nu}^a - \partial_{\nu} W_{\mu}^a
                + g\epsilon^{abc} W_{\mu}^b W_{\nu}^c. \nonumber
\end{eqnarray}
$ \sigma^a $ are the Pauli spin matrices, $W_{\mu}^a $ are the $SU(2) $
gauge fields, $B_{\mu} $ are the $U(1)$ gauge fields and $\phi $ is the
Higgs doublet. The $ SU(2)_L\times U(1)_Y $ symmetry of this lagrangian
is broken to the $ U(1)_{em} $ symmetry of electromagnetism by the
Higgs field acquiring a vacuum expectation value
$ \phi^{\dagger} = (0, \eta /\sqrt{2} ) $. With the standard field basis
\begin{eqnarray}
Z_{\mu} = \cos \theta_{W} W_{\mu}^3 - \sin \theta_{W} B_{\mu} &   &
A_{\mu} = \sin \theta_{W} W_{\mu}^3 + \cos \theta_{W}
B_{\mu} \nonumber \\
W_{\mu}^{+} = \frac{1}{\sqrt{2}} (W_{\mu}^1 -iW_{\mu}^2 )&   &
W_{\mu}^{-} = \frac{1}{\sqrt{2}} (W_{\mu}^1 +iW_{\mu}^2 ) \nonumber
\end{eqnarray}
the boson masses generated by the Higgs mechanism are
$$
M_{W} = \frac{1}{2} g\eta, \hspace{10mm} M_{Z} = \frac{1}{2}g_z \eta,
\hspace{10mm} M_{H} = \sqrt{2\lambda \eta^2}.
$$
The Weinberg angle $ \theta_{W} $ is given by
$ \tan \theta_{W} = g^{\prime} /g $, and
$ g_z = \sqrt{g^2 + {g^{\prime}}^2 } $.

Consider time independent field configurations with
$ \phi_{u}=W_{\mu}^{+}=W_{\mu}^{-}=A_{\mu}=0 $ and $ Z_0 =0 $,
then the energy functional is given by
$$
E=\int d^3 x \left( \frac{1}{4} Z_{ij} Z_{ij}
+ |(\nabla_{i} + \frac{ig_z}{2}Z_{i})\phi_d |^2
+ \lambda (|\phi_{d} |^2 -\eta^2 /2 )^2 \right),
$$
where
$$
Z_{ij} = \nabla_i Z_j -\nabla_j Z_i.
$$
This is the same as for the Abelian-Higgs model and so there is a
Nielsen-Olesen string configuration given by
$$
\phi_d = \frac{\eta}{\surd{2}} f(r)e^{iN\theta } ,\hspace{15mm}
 Z_{r}=0, \hspace{15mm}
 Z_{\theta}=-\frac{2N}{g_z r} a(r),
$$
which extremises the energy. The functions $ f(r) $ and $ a(r) $
are identical to those found by Nielsen and Olesen \cite{NO}.

Consider a general perturbation to the solution of the form
\begin{eqnarray}
\phi_u & = & \epsilon \delta \phi_u e^{-i\omega t} \hspace{21mm}
\phi_u^{\ast} =\epsilon \delta \phi_u^{\ast}
e^{i\omega t} \nonumber \\
\phi_d & = & \phi_d^c + \epsilon \delta \phi_d e^{-i\omega t}
\hspace{12mm} \phi_d^{\ast} ={\phi_d^c}^{\ast} + \epsilon \delta
\phi_d^{\ast} e^{i\omega t} \nonumber \\
W_{\mu}^{+} & = & \epsilon \delta W_{\mu}^{+} e^{-i\omega t}
\hspace{16mm}
W_{\mu}^{-} =\epsilon \delta W_{\mu}^{-} e^{-i\omega t} \nonumber \\
Z_{\mu} & = & Z_{\mu}^c + \epsilon \delta Z_{\mu} e^{-i\omega t}
\hspace{10mm} A_{\mu} = \epsilon \delta A_{\mu} e^{-i\omega t} \nonumber.
\end{eqnarray}
This gives corrections to the action
$$
S=S(\phi_d^c ,Z_{\mu}^c ) + \epsilon^2 \frac{1}{2}
\int d^4 x \delta \chi^{\dagger} {\cal D} \delta \chi + O(\epsilon^3),
$$
where ${\cal D}$ is the perturbation operator and
$$
\delta \chi^{\dagger} = (\delta\phi_u e^{-i\omega t},
\delta\phi_u^\ast e^{i\omega t},
\delta W_{\mu}^+ e^{-i\omega t},
\delta W_{\mu}^- e^{-i\omega t},
\delta\phi_d e^{-i\omega t},
\delta\phi_d^\ast e^{i\omega t},
\delta Z_{\mu} e^{-i\omega t},
\delta A_{\mu} e^{-i\omega t}).
$$
The equations of motion for the perturbations
$ {\cal D} \delta \chi =0 $,
could in principle be coupled equations in twenty fields. For the
string background however, they separate into four sets of coupled
equations. The first two are of the form
\begin{eqnarray}
{\cal D}_1 \left( \begin{array}{c}
              \delta \phi_u \\ \delta W_{\mu}^+
        \end{array}
   \right) = 0,  \hspace{20mm}
{\cal D}_2 \left( \begin{array}{c}
              \delta \phi_u^\ast \\ \delta W_{\mu}^-
        \end{array}
   \right) = 0,
\end{eqnarray}
which, since the string solution is invariant under charge conjugation,
are just charge conjugates, so we need only consider the first of these.
The third is of the form
$$
{\cal D}_3\left( \begin{array}{c}
              \delta \phi_d \\ \delta \phi_d^\ast \\ \delta Z_{\mu}
        \end{array}
   \right) = 0 ,
$$
which contains only the fields of the Nielsen-Olesen vortex, and was
addressed in \cite{MGMH}. The fourth equation is
$$
{\cal D}_4 \delta A_{\mu} = (\eta^{\mu\nu} \partial^2
 -\partial^{\nu} \partial^{\mu} )\delta A_{\mu} =0,
$$
which is just the usual wave equation of electromagnetism. The last two
equations do not contain any modes which can decrease the winding number,
and so we must look at the solutions of (1) above for decay modes.
As mentioned in \cite{MGMH} these equations contain linear derivative
terms and gauge degrees of freedom, both of which can be removed by
choosing the background gauge conditions
\begin{eqnarray}
F_1 (W_{\mu}^+ ) & = & \partial^{\mu}W_{\mu}^+
-ig\cos\theta_W Z_c^{\mu}W_{\mu}^+
-\frac{ig}{\sqrt{2}} {\phi_d^c}^\ast \phi_u = 0 \\
F_2 (W_{\mu}^- ) & = & \partial^{\mu}W_{\mu}^-
+ig\cos\theta_W Z_c^{\mu}W_{\mu}^-
+\frac{ig}{\sqrt{2}} \phi_d^c \phi_u^\ast = 0 \\
F_3 (Z_{\mu} ) & = & \partial^{\mu}Z_{\mu} - \frac{ig_z}{2}(\phi_d^\ast
\phi_d^c - {\phi_d^c}^\ast \phi_d ) = 0 \\
F_4 (A_{\mu} ) & = & \partial^{\mu} A_{\mu} = 0
\end{eqnarray}
which are imposed by adding the gauge fixing terms
$$
{\cal L_{GF} }= \frac{1}{2} \sum_{i=1}^{4} |F_i |^2
$$
to the lagrangian. This enables us to separate out the linear time
derivatives and to set up gauge-fixed eigenvalue equations, which
for equation (1) above are
\begin{eqnarray}
M^{GF} \left( \begin{array}{c}
           \delta \phi_u \\ \delta W_i^+
           \end{array}
           \right) = \omega^2 \left(
           \begin{array}{c}
           \delta \phi_u \\ \delta W_i^+
           \end{array}
           \right)
\end{eqnarray}
and
\begin{eqnarray}
(- \nabla^2 + 2ig\cos\theta_W Z_k^c \nabla_k
+g^2 \cos^2 \theta_W Z_k^c Z_k^c
+\frac{g^2}{2} |\phi_d^c |^2 )\left(
   \begin{array}{c}
      \delta W_0^+ \\ \delta W_z^+
   \end{array}
   \right) = \omega^2 \left(
   \begin{array}{c}
      \delta W_0^+ \\ \delta W_z^+
   \end{array}
   \right),
\end{eqnarray}
where $i=1,2$, and the gauge fixed perturbation operator is given by
$$
M^{GF}=\left( \begin{array}{cc}
         D_1 & X_i \\
         X^{\ast}_j & D_{2_{ji}}
       \end{array}
       \right)
$$
where
\begin{eqnarray}
D_1 & = &-\nabla^2 +ig_z\cos 2\theta_W Z_k^c\nabla_k + \frac{g_z^2}{4}
\cos^2 2\theta_W Z_k^c Z_k^c + 2\lambda (|\phi_d^c |^2 -\eta^2 /2)
+\frac{g^2}{2} |\phi_d^c |^2 \nonumber \\
D_{2_{ji}} & = & \delta_{ji}(-\nabla^2 +2ig\cos\theta_W Z_k^c\nabla_k
+ g^2 \cos^2 \theta_W Z_k^c Z_k^c +\frac{g^2}{2} |\phi_d^c |^2)
+2ig\cos\theta_W Z_{ji}^c \nonumber \\
X_i & = & ig\sqrt{2} (\nabla_i +\frac{ig_z}{2}Z_i^c)
\phi_d^c \nonumber \\
X_j^\ast & = &-ig\sqrt{2} (\nabla_j -\frac{ig_z}{2}Z_j^c)
{\phi^c_d}^\ast .\nonumber
\end{eqnarray}
The $\delta W_0^+$ and $\delta W_z^+$ perturbations decouple because
the background string solution is independent of $t$ and $z$.

The gauge condition used means that the gauge fixing terms must be
accompanied by Fadeev-Popov ghost terms. This results in an
accompanying eigenvalue equation for the ghost field $\Lambda^+$
\begin{eqnarray}
(- \nabla^2 + 2ig\cos\theta_W Z_k^c \nabla_k +g^2 \cos^2
\theta_W Z_k^c Z_k^c
+\frac{g^2}{2} |\phi_d^c |^2 ) \Lambda^+ = \omega^2 \Lambda^+.
\end{eqnarray}
This is the same equation as (7) and so
$\Lambda^+$, $\delta W_0^+$ and $\delta W_z^+$
all have the same eigenvalue spectra. Of the total of 5 eigenmodes
in (6), (7) and (8), only three should be physical, corresponding
to the three spin states of the massive $W$ boson. This is ensured
by the ghosts canceling one linear combination of $\delta W_0^+$ and
$\delta W_z^+$, and one of the eigenmodes of $M_{GF}$.

It may seem that the gauge choice (2)-(5) introduces unnecesary
complications and a difficulty in identifying the physical modes.
However, we are looking for decay modes where $\omega^2 <  0$ and
the eigenvalues of (8) are positive, so there is no difficulty in
identifying the field configurations of the decay modes. It is also
worth pointing out that (2)-(5) results in a considerable
simplication in the resulting eigenvalue equations (as opposed to
the temporal gauge for instance).

We expand the scalar and gauge fields in total angular momentum states.
For the gauge fields they are
$$
W^+_{\uparrow} =\frac{e^{-i\theta}}{\sqrt{2}}
(W^+_r -\frac{i}{r}W^+_{\theta} ),
\hspace{8mm} \mbox{and} \hspace{8mm}
W^+_{\downarrow} =\frac{e^{i\theta}}{\sqrt{2}}
(W^+_r +\frac{i}{r}W^+_{\theta} ).
$$
The total angular momentum operator for the gauge fields is
$J_z=L_z+S_z$ where
$$
L_z= -i\frac{d}{d\theta}, \hspace{10mm} \mbox{and} \hspace{10mm}
(S_zW^+)_j = -i\epsilon_{3jk} W^+_k,
$$
with $(S_zW^+)_{\uparrow} = +W^+_{\uparrow}$, and
$(S_zW^+)_{\downarrow}=-W^+_{\downarrow}$. So the suffices $(\uparrow )$
and $(\downarrow )$ identify gauge fields with spin up and spin down
respectively. The background string solution in total angular momentum
states is
$$
\phi_d^c = \frac{\eta}{\sqrt{2}}f(r)e^{iN\theta}, \hspace{8mm}
Z_{\uparrow}^c=\frac{i\sqrt{2}N}{g_z r}a(r)e^{-i\theta}, \hspace{8mm}
Z_{\downarrow}^c=\frac{-i\sqrt{2}N}{g_z r}a(r)e^{i\theta},
$$
and the perturbations are
$$
\delta\phi_u = \sum_{m^\prime} s_{m^\prime} e^{im^\prime \theta},
\hspace{10mm}
\delta W_{\uparrow}^+ = \sum_m -iw_m e^{i(m-1)\theta}, \hspace{10mm}
\delta W_{\downarrow}^+ = \sum_m iw_{-m}^{\ast} e^{i(m+1)\theta},
$$
with $N+m=m^{\prime}$. It is useful to rescale
$$
\phi=\frac{\eta}{\sqrt{2}}\phi,\hspace{5mm}
Z_i=\frac{\eta}{\sqrt{2}}Z_i,
\hspace{5mm}W^+_i=\frac{\eta}{\sqrt{2}}W^+_i,
\hspace{5mm}W^-_i=\frac{\eta}{\sqrt{2}}W^-_i,
\hspace{5mm}r=\frac{2\sqrt{2}}{g_z \eta}\rho ,
$$
to give dimensionless variables. The eigenvalues are now in units $g_z^2
\eta^2/8$ and the magnetic field strength is in units $g_z\eta^2/4 =
m_Z^2 /g_z$. Substituting in the above gives the eigenvalue equations
\begin{eqnarray}
\left( \begin{array}{ccc}
       D_1 & A & B \\
       A & D_2 & 0 \\
       B & 0 & D_3
       \end{array} \right)
       \left( \begin{array}{c}
        s_{m^\prime} \\ w_m \\ w_{-m}^{\ast}
        \end{array} \right) & = & \omega^2 \left(
        \begin{array}{c}
         s_{m^\prime} \\ w_m \\ w_{-m}^{\ast}
        \end{array} \right)
\end{eqnarray}
where
\begin{eqnarray}
D_1 & = & -\nabla_\rho^2 + \frac{(m^{\prime} +aN\cos^2\theta_W)^2}
{\rho^2}+\beta (f^2 -1) +2f^2 \cos^2 \theta_W \nonumber \\
D_2 & = & -\nabla_\rho^2 + \frac{((m-1) +2aN\cos^2\theta_W)^2}{\rho^2}
+2f^2 \cos^2 \theta_W + 4\cos^2 \theta_W \frac{N}{\rho}\frac{da}{d\rho}
\nonumber \\
D_3 & = & -\nabla_\rho^2 + \frac{((m+1) +2aN\cos^2\theta_W)^2}{\rho^2}
+2f^2 \cos^2 \theta_W - 4\cos^2 \theta_W \frac{N}{\rho}\frac{da}{d\rho}
\nonumber \\
A & = & 2\left( \nabla_\rho f -\frac{Nf}{\rho}(1-a) \right) \nonumber \\
B & = & -2\left( \nabla_\rho f +\frac{Nf}{\rho}(1-a) \right) \nonumber
\end{eqnarray}
and $\beta =8\lambda /g_z^2$.

If we resolve $s_{m^\prime},w_m,w_{-m}^{\ast}$ into real and imaginary
parts, the complex eigenvalue problem seperates into two eigenvalue
problems with explicitly real fields
$$
\left( \begin{array}{ccc}
     D_1 & A & B \\
     A & D_2 & 0 \\
     B & 0 & D_3
     \end{array}  \right) \left(
     \begin{array}{c}
       s^r_{m^\prime} \\ w_m^r \\ w^r_{-m}
              \end{array} \right) = \omega^2 \left(
     \begin{array}{c}
       s^r_{m^\prime} \\ w_m^r \\ w^r_{-m}
       \end{array} \right)
$$
and
$$
\left( \begin{array}{ccc}
     D_1 & A & B \\
     A & D_2 & 0 \\
     B & 0 & D_3
     \end{array}  \right) \left(
     \begin{array}{c}
       s^i_{m^\prime} \\ w_m^i \\ -w^i_{-m}
       \end{array} \right) = \omega^2 \left(
     \begin{array}{c}
       s^i_{m^\prime} \\ w_m^i \\ -w^i_{-m}
       \end{array} \right)
$$
where $D_\gamma,A$ and $B$ are given above.
The profiles for the Nielsen-Olesen vortex were solved for with a
relaxation method on the energy functional, and then substituted into
the eigenvalue equations above. This eigenvalue problem was solved
for a range of $\beta$ and $\theta_W$ with the boundary
conditions $s_{m^\prime},w_m,w_{-m} \rightarrow 0$ as $\rho
\rightarrow \infty$ by the standard matrix packages incorporated
into {\sc matlab}.

For $1>\beta>0.04$ a linear discretisation of $\rho$ was used. For
$\beta<0.04$ the scales of the  Higgs core and the gauge field core
are too different to use this linear descretisation
so we followed Yaffe \cite{Yaf} in using the convenient transformation
$$
\xi = \frac{1}{\ln (m_H/m_Z)} \ln \left( \frac{1+m_Hr}{1+m_Zr} \right).
$$
This maps the line $0\leq r <\infty$ to $0\leq \xi <1$.
Linear discretisation of $\xi$ results in more points in the core
region and is better suited to studying very small and very large values
of $\beta =m_H/m_Z$.
With this map we were able to look at $\beta$ down to $1\times 10^{-4}$.

The resulting stability lines are shown in the Figure. All the lines
for the various decay modes separate the region on the left hand side
where the string is unstable, from the region on the right hand side
where the string is stable with respect to a particular decay mode.

The two decay modes of Ach\'ucarro {\it et al} correspond to
$m^{\prime} =0$ and $m= -1$.
For the $m^\prime =0$ mode $\phi_u$ acquires a
non-zero value at the core of the string, and then there is nothing to
stop the string from completely unwinding and decaying to the vacuum.
For the $m =-1$ mode $W^+_{\downarrow}$ acquires a non-zero value at the
core of the string, where its magnetic moment is aligned
with the magnetic field of the string.
The term giving rise to this $W$-condensation is
$-4\cos^2\theta_W\delta W^+_{\downarrow}Na^\prime/\rho$,
which is of the generic form
$-{\bf m.B}$ for the interaction between a magnetic moment ${\bf m}$
and a magnetic field ${\bf B}$. This term comes from the non-abelian nature
of the field tensor.
Ach\'ucarro {\it et al} showed that a $W$-condensed string of winding
number $N$ has the same energy as a string of winding $N-1$, and
so the $m =-1$ mode corresponds to a unit unwinding. It is this
$m=-1$ mode which is refered to as $W$-condensation in \cite{Perk}
and \cite{Achu}. In the background gauge, however, far
from the string $\phi_u$ can readily be indentified as the longitudinal
$W$ gauge field, and so all the modes we have found can be called
$W$-condensation.

For the $N=1$ string the relation $N+m=m^{\prime}$ gives that the
$m^{\prime} =0$ and $m=-1$ modes are one and the same with both
$\phi_u$ and $W^+_{\downarrow}$ acquring non-zero values
at the core of the string. The upper part of the stability line in
the Figure agrees with that found by James {\it et al}. This is
different from the stability line in \cite{Perk}
($\sqrt{\beta} =4\cos \theta_W $) which
was found analytically by using fairly crude approximations
for the string solution and for the profiles of the $W$-fields.
A large part of this difference could be that the decay mode
of the $N=1$ string has non-trivial field configurations for
the $W$-fields and for $\phi_u$ as well, whereas the analysis
in \cite{Perk} contained no $\phi_u$ field.

For the $N=2$ string as well as the two decay modes mentioned above
there is a third decay mode for which $m^{\prime} =-1$, $m=-3$. We
conjecture that this decay mode corresponds to the $N=2$ string
decaying to a string of winding number $N=-1$. Since the $Z$-strings are
non-topological and an $N=-1$ string has the same energy as an $N=1$
string, such a decay is to be expected, albeit with a larger energy
barrier than the decay to an $N=1$ string. This later point is
consistent with the $m^{\prime} =-1$ decay mode having a much larger
region of stability in the parameter space than the decay mode to
an $N=1$ string.

Generalising this to strings of winding number $N$, means that we
would expect there to be $2N-1$ decay modes corresponding to the string
decaying to strings of winding number $N-1, \ldots ,-N+1$. Furthermore
we would expect that the stability lines for these decays to be such
that those for larger units of unwinding would have greater regions of
stability, due to the increasing energy barrier for such decays.

The stability lines for the $N=3$ string, appear
to be consistent with this interpretation (as are those for $N=1$ and
$N=2$ strings). It is
also consistent with the result of MacDowell and T\"ornkvist that
strings of winding number $N$ have regions in the parameter space for which
they are unstable to the formation of $W$-fields with angular momentum
$m$ such that $ -2N < m <0$.

For the strings we studied ($N=1,2,3$), all have a stability line
close to
the $\sqrt{\beta}$ axis for the mode with $m=-2N$. For all the strings
this line is in approximately the same place ($\sin^2 \theta_W \simeq
0.02$) and approaches the
$\sqrt{\beta}$ axis as the number of points in the lattice is increased.
We believe that in the continuum limit this $m=-2N$ mode will have zero
eigenvalue along the line $\sin^2 \theta_W =0$, and the deviation
of the line from this value for finite lattices (100 points) is an
indication of the error.

For the complex conjugate problem with the fields $\phi_u^\ast$ and
$W^-$,the stability lines are of course the same. The `$W$-condensation'
mode is where $W^-_{\uparrow}$ is non-zero at the core of the string,
so the magnetic
moment is still aligned with the magnetic field of the string.

It must be noted that all the modes involve
non-trivial field configurations for $\phi_u$,$W^+_{\uparrow}$ and
$W^+_{\downarrow}$.

Finally, we consider the large $N$ limit. If we assume that the
$\phi_u$ and $W$ perturbations drop to zero whilst inside
the core, we need only consider the small $\rho$ behaviour of
the string solution, which is
$f \rightarrow f_N \rho^N$, $a \rightarrow \rho^2 /2$ as $\rho
\rightarrow 0$, where $f_N$ is a constant.
Substituting this into the perturbation equation (9)
gives
\begin{eqnarray}
D_1 & = & -\nabla_\rho^2
+ \frac{(m^{\prime} +\rho^2N\cos 2\theta_W/2)^2}{\rho^2}
+\beta (f_N^2\rho^{2N} -1) +2f_N^2\rho^{2N} \cos^2 \theta_W \nonumber \\
D_2 & = & -\nabla_\rho^2
+ \frac{((m-1) +\rho^2N\cos^2\theta_W)^2}{\rho^2}
+2f_N^2\rho^{2N} \cos^2 \theta_W + 4N\cos^2 \theta_W  \nonumber \\
D_3 & = & -\nabla_\rho^2
+ \frac{((m+1) +\rho^2N\cos^2\theta_W)^2}{\rho^2}
+2f_N^2\rho^{2N} \cos^2 \theta_W - 4N\cos^2 \theta_W  \nonumber \\
A & = & f_N N\rho^{N+1} \nonumber \\
B & = & f_N N\rho^{N+1} -4f_N N\rho^{N-1}. \nonumber
\end{eqnarray}
So for large $N$, the perturbations in $\phi_u$ and the $W$ fields
decouple.
Considering terms up to $\rho^2$ it was shown in \cite{Achu} that the
equation for $\phi_u$ for $m^{\prime}=0$ has a stability line given by
$\sqrt{\beta} = |1-2\sin^2 \theta_W |$. The Figure seems to indicate
that the $m^{\prime} =0$ line is approaching this limit but there is
no stability region in the bottom left of the plot, which is not
altogether suprising since $N=3$ is not large $N$.

Lastly, we present an explanation for the movement of the $m=-1$
stabilitiy line towards the line $\sin^2 \theta_W =1$ for
increasing winding number.
Consider the equation for $W^+_{\downarrow}$ with $m=-1$ for large $N$,
with the assumption that the perturbation drops rapidly to zero
within the core, so that we need keep terms only up to $O( \rho^2 )$:
$$
\left( -\frac{d^2}{d\rho^2} -\frac{1}{\rho}\frac{d}{d\rho}
+N^2\rho^2\cos^4\theta_W -4N\cos^2\theta_W \right)w^r_1 =
\omega^2 w^r_1.
$$
This has a solution $w^r_1 = c\exp (-N\cos^2\theta_W\rho^2 /2)$
for which $\omega^2 = -2N\cos^2 \theta_W$, and so large $N$ strings
are unstable to `$W$-condensation' for all values of the parameters.
This is consistent with the stability regions of the $m=-1$ modes
displayed in the Figure.

\subsection*{Acknowlodgements}
We are grateful to Ana Ach\'ucarro for helpful conversations.
This work was supported by PPARC: MG by studentship number
93300941 and MH by Advanced Fellowship number B/93/AF/1642.

\newpage

\section*{Figure captions}

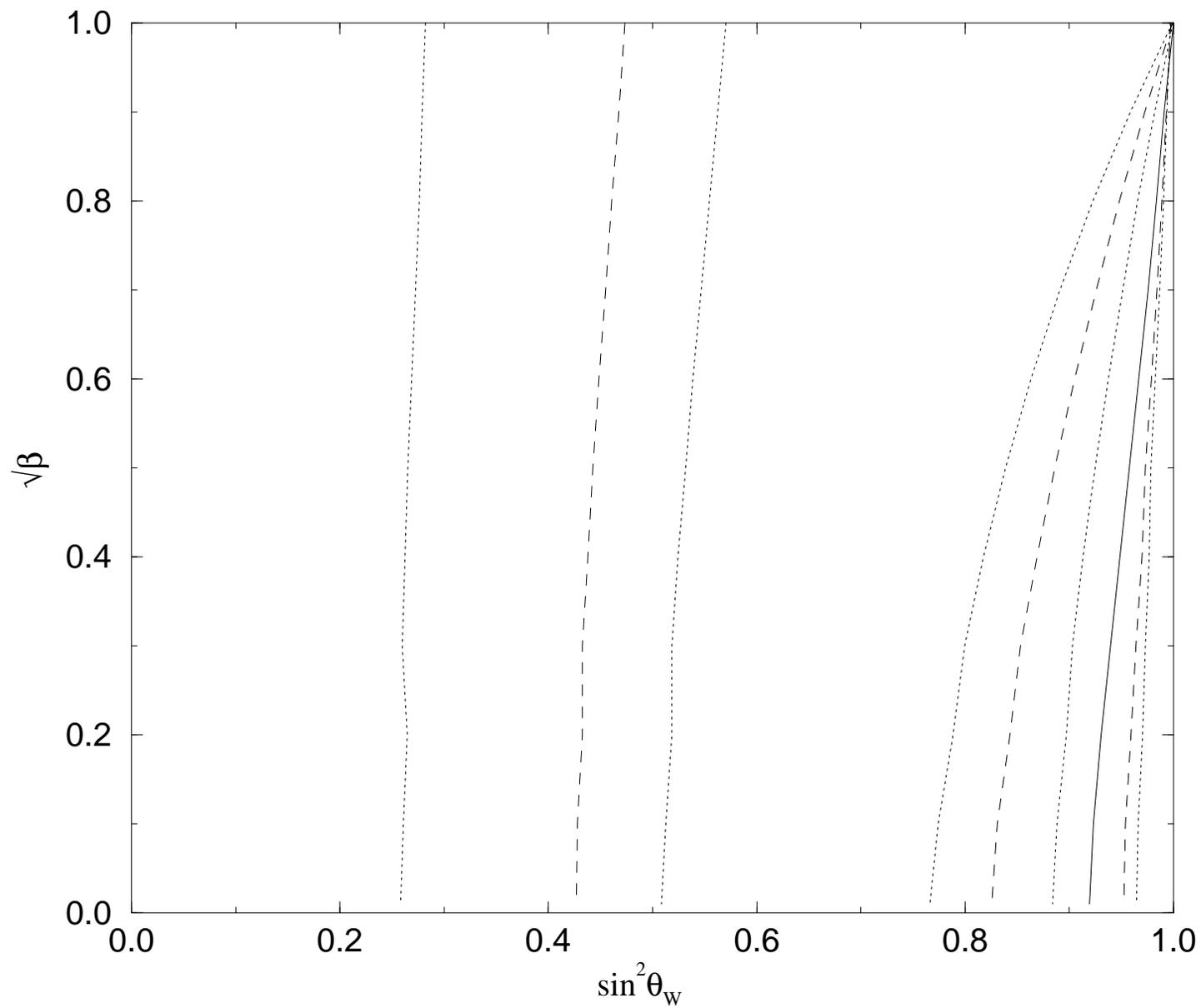
{\bf Figure 1:} The stability lines for a) the $N=1$ string (solid line),
b) the $N=2$ string (dashed line) and c) the $N=3$ string (dotted line).
The angular momentum $m$ of the modes are (reading from right to left)
a) $m=-1$, b) $m=-1,-2,-3$, c) $m=-1,-2,-3,-4,-5$.


\begin{thebibliography}{99}
\bibitem{V&A} T. Vachaspati and A. Ach\'ucarro, {\it Phys.~Rev.} D44,
3067 (1991)
\bibitem{Hind} M. Hindmarsh, {\it Phys. Rev. Lett.} 68, 1263 (1991)
\bibitem{Vach} T. Vachspati, {\it Phys. Rev. Lett.} 68, 1977 (1991)
\bibitem{T&B} T. Vachspati and M. Barriola, {\it Phys. Rev. Lett.} 69,
1867 (1992)
\bibitem{James} M. James, L. Perivolaropoulos and T. Vachaspati,
{\it Nucl. Phys.} B395, 534 (1993)
\bibitem{Perk} W. B. Perkins, {\it Phys.~Rev.} D47, 5224 (1993)
\bibitem{Achu} A. Ach\'ucarro, R. Gregory, J. A. Harvey and K. Kuijken,
{\it Phys. Rev. Lett.} 72, 3646 (1994)
\bibitem{D&T} S. W. MacDowell and O. T\"ornkvist, NORDITA-94/52 P
\bibitem{NO} H. Nielsen and P. Olesen, {\it Nucl. Phys.} B61, 45 (1973)
\bibitem{MGMH} M. Goodband and M. Hindmarsh, hep-ph/9503457
\bibitem{Yaf} L. G. Yaffe {\it Phys.~Rev.} D40, 3463 (1989)
\bibitem{A&O} J. Ambj{\o}rn and P. Olesen, {\it J. Mod. Phys.} A5, 4525
(1990)
\bibitem{T&T} D. R. Tilley and J. Tilley, {\it Superfluidity and
Superconductivity $2^{nd}$ ed. } (Adam Hilger Ltd, Bristol and Boston,
(1986)

\end{thebibliography}
\end{document}